\documentclass[12pt,a4paper,superscriptaddress]{revtex4}
\usepackage{graphicx}
\usepackage{graphics}
\usepackage{amsmath}
\usepackage{dcolumn}
\usepackage{amssymb}
\usepackage{bm}
\usepackage[latin1]{inputenc}
\newcommand{\be}{\begin{equation}}
\newcommand{\ee}{\end{equation}}
\newcommand{\bea}{\begin{eqnarray}}
\newcommand{\eea}{\end{eqnarray}}

\newcommand{\pa}{\partial}

\begin{document}

\title{\Large 
On the decoupling of heavy states in higher-derivative supersymmetric field theories\\
 }

\author{\large M. Cveti\v{c}}
\email{cvetic@physics.upenn.edu}
\affiliation{Department of Physics and Astronomy, University of Pennsylvania,\\
Philadelphia, PA 19104-6396, USA}
\affiliation{
 Center for Applied Mathematics and Theoretical Physics,\\
 University of Maribor, SI2000 Maribor, Slovenia}
\author{\large T. Mariz}
\email{tmariz@fis.ufal.br}
\affiliation{Instituto de F\'\i sica, Universidade Federal de Alagoas,\\ 
57072-270, Macei\'o, Alagoas, Brazil}
\author{\large A. Yu. Petrov}
\email{petrov@fisica.ufpb.br}
\affiliation{Departamento de F\'{\i}sica, Universidade Federal da Para\'{\i}ba,\\
 Caixa Postal 5008, 58051-970, Jo\~ao Pessoa, Para\'{\i}ba, Brazil}

\begin{abstract}
\medskip
\medskip
\medskip
{\center{ \large \bf Abstract:}}\\
{\noindent {We study the problem of decoupling of heavy chiral superfields  in four-dimensional  $N=1$ supersymmetric field theories with Lorentz-invariant and Lorentz-violating higher-derivative terms.
 We demonstrate that the earlier found effect of large logarithmic quantum corrections, due to  heavy chiral superfields, takes place not only if the theory possesses quantum divergences, but also for essentially finite theories involving higher derivative terms, both Lorentz-invariant and Lorentz-breaking  ones.}}
\end{abstract}

\maketitle
\newpage
\section{Introduction}

The four-dimensional  $N=1$ supersymmetric effective action with chiral superfields naturally arises as a subsector of  the low-energy limit of  compactified superstring theory.  
In this context it is important to address  the impacts of massive modes, associated with the string theory scale $M_{String}$, on the effective four-dimensional action. In perturbative  heterotic string theory  the string scale  is determined to be  $M_{String}\sim 10^{-2} M_{Planck}$ \cite{CEW}. For  perturbative  Type II string theory  compactifications  the string scale could, in principle, be pushed all the way  to the electroweak scale \cite{Antoniadis,Nima}.
 In the effective field theory,  the impacts of massive modes are subject to 
the decoupling theorem \cite{Appel}.  Namely,  if one considers the theory involving light 
and heavy (super)fields,  say, with mass $M$  which is  of the order of $M_{String}$, the effective action of light fields   is represented  as a sum of dimension four-operators,  and terms  suppressed by the factors proportional to  $M^{-n}$ (with $n\ge1$). The latter ones decouple, as $M\to \infty$.  For sectors of  $N=1$ supersymmetric theory with chiral superfields,  this conclusion has been verified at the tree level in \cite{CEW}. When  quantum corrections are taken into account, it turns to be that there are new corrections, which are not suppressed at large $M$,  but instead their contribution grows logarithmically with  $M$  \cite{BCP}.  This result formally does not contradict the Appelquist-Carazzone theorem  \cite{Appel}, as after  an appropriate definition of physically measured couplings,  the effective action involves light superfields only, whose  quantum corrections are not suppressed  \cite{BCP}. 

The  $N=1$ supersymmetric field theory considered in \cite{BCP} had the following features: first, it  did not not involve higher derivatives and  second, at the quantum level it was  divergent, so that quantum corrections depended on the normalization scale $\mu$, involving terms proportional to  $\ln\frac{M^2}{\mu^2}$, which were responsible for the above-mentioned significant quantum corrections.  In this paper we would like to address  how the decoupling theorem manifests itself in higher-derivative (super)field theories. Note that higher-derivative terms naturally emerge in effective theories  of compactified string theory, cf.~\cite{HDString}.  In this paper, we focus on the 
the supersymmetric higher derivative sector  with chiral superfields, and calculate quantum corrections there.
We employ the superfield approach to calculate the effective potential described in great details in \cite{BK0}, and follow the conventions of that book. Technically, we use the methodology of a summation over cycle-like one-loop diagrams developed for superfield theories in \cite{YMEP} and further applied to  different theories in many papers, including \cite{BCP}. We consider several examples of higher derivative terms,  including those that  break  Lorentz invariance. 
The case with  Lorentz-violating  terms  is of special interest:  as it was argued in \cite{CMR},  the presence of Lorentz-violating terms, which are of a special form $(n\cdot\partial)^N$, with $n_a$ is a space-like vector, and $N$ is a positive integer umber, allows for  an implementation of  higher derivatives without ghosts. In this paper, such terms  are implemented in a superfield context for the first time, thus developing a methodology for Lorentz-violating extension to superfield models, earlier proposed in \cite{LVWZ}. In the superfield formalism, these Lorenz-violating terms are added
 to a classical action in a manifestly supersymmetric way, i.e. without deformations of supersymmetry algebra or  an introduction of new superfields.

Our key result is  that the presence of higher derivatives in an effective action introduces  large, typically logarithmic, quantum corrections, independent of the fact whether the theory is finite or divergent.

The structure of the paper is the following. In  section 2, we  summarize the results in  $N=1$ supersymmetric theories with massive superfields  without higher derivatives,  where the renormalization 
gives rise to large logarithmic quantum corrections. In the section 3, we consider  $N=1$ supersymmetric theories with different examples of higher-derivative terms, which are superficially finite, and derive how large quantum corrections arise in such theories. In  section 4  we summarize the  results and discuss implications.

\section{Quantum corrections in theories without higher derivatives}

We start our study by considering  the four-dimensional $N=1$ supersymmetric theory with chiral superfield, without higher derivatives. The simplest superfield model involving both light (massless) and heavy superfields is given by the classical action
\bea
\label{s}
S=\int d^8z (\phi\bar{\phi}+\Phi\bar{\Phi})+ \Big(\int d^6z (\frac{1}{2}(M\Phi^2+\lambda\Phi\phi^2+f\phi\Phi^2)+\frac{g}{3!}\phi^3)+h.c.\Big).
\eea
 Here $\phi$ is a light  chiral superfield, which for the sake of simplicity, we choose it to be a massless chiral superfield.
$\Phi$ is a heavy superfield with a  large mass $M$. As a concrete example we choose  $M$ to be of the order of $M_{String}\sim 10^{-2}M_{Planck}$ in perturbative heterotic string theory compactification, as in \cite{CEW,BCP}.
Note that in perturbative Type II string compactification $M_{String}$ could be chosen to be many orders of magnitude smaller than $M_{Planck}$. 
 
The above  tree-level action slightly differs from the one considered in \cite{CEW}, and in \cite{BCP},
where the vertex proportional to $f$ was absent. We introduced it here to couple heavy and light superfields in a manner convenient for our study. Nevertheless, the tree-level situation is not qualitatively  different from \cite{CEW}. Indeed, one can find the equation of motion for $\Phi$:
\bea
-\frac{1}{4}\bar{D}^2\bar{\Phi}+(M+f\phi)\Phi+\frac{\lambda}{2}\phi^2=0,
\eea
so that we can write the solution for $\Phi$ via an iterative method as $\Phi=\Phi_0+\Phi_1+\ldots$, where the zero approximation is
\bea
\Phi_0=-\frac{\lambda\phi^2}{2(M+f\phi)},
\eea 
being of first order in $\frac{1}{M}$, and for $k$-th order one has
\bea
\Phi_{k+1}=\frac{1}{4}\frac{1}{M+f\phi}\bar{D}^2\bar{\Phi}_k,
\eea
i.e. the $k$-th order is suppressed at least as $M^{-(k+1)}$. The same situation occurred within the study of the tree-level effective action in \cite{CEW}.

However, the situation turns out to be much more delicate if we consider quantum corrections. Although we restrict ourselves to the one-loop order, the results are remarkable. In this case it is very easy to illustrate the origin of significant quantum corrections.

For the calculation of the one-loop corrections, we employ the loop expansion formalism.  To do this,  we split the superfields $\{\phi,\Phi\}$ 
into a sum of the background (classical) superfields $\{\phi_0,\Phi_0\}$ and the quantum ones $\{\phi_q,\Phi_q\}$, via the rule $\phi\to\phi_0+\phi_q$, and  $\Phi\to\Phi_0+\Phi_q$.
It is well known that within the one-loop approximation, we must keep only the second order in quantum superfields over which we should then integrate.

For the sake of simplicity, we choose that the light field $\phi$ is a purely a background one, while the heavy one $\Phi$ is a purely quantum one. (Indeed, as we already noted, the contributions that arise due to the presence of background heavy fields yield only corrections suppressed by  $M^{-n}$,  with $n\geq 1$, cf. \cite{BCP}, so, we can neglect the background $\Phi$ within the lower-order approximation.) The quadratic action of the quantum field $\Phi_q=\Phi$  takes the form
\bea
S_q=\int d^8z \Phi\bar{\Phi}+ \Big(\frac{1}{2}\int d^6z (M+f\phi)\Phi^2+h.c.\Big)\, .
\eea

We incorporate the mass into the background field $\Psi\equiv M+f\phi$. As a result, the propagator of $\Phi$ has the usual form \cite{BK0}:
$$
<\Phi\bar{\Phi}>=-\frac{\bar{D}^2 D^2}{16\Box}\delta^8(z_1-z_2).
$$

Therefore, the one-loop effective potential is contributed by a sum of supergraphs depicted in Fig.~1. 

\vspace*{2mm}

\begin{figure}[!h]
\begin{center}
\includegraphics[angle=0,scale=1.00]{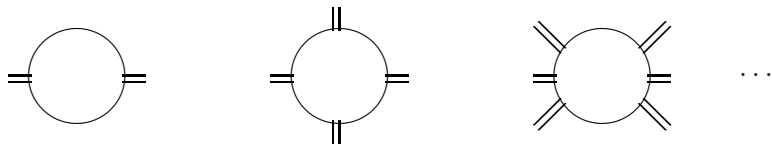}
\end{center}
\caption{The double legs denote alternating $\Psi$ and $\bar{\Psi}$ background fields. The single internal line denotes $<\Phi\bar{\Phi}>$ propagators.}
\end{figure}

\vspace*{2mm}

We note that this set of supergraphs completely describes the one-loop effective potential in a  generic chiral superfield theory whose quadratic action is
\bea
\label{gentheory}
S=\int d^8z \Phi \hat{T}\bar{\Phi}+\frac{1}{2}(\int d^6z \Psi \Phi^2+h.c.)\, ,
\eea
where $\Psi$ is any background chiral superfield, and $\Phi$ is a quantum one, and $\hat{T}$ is any operator commuting with supercovariant derivatives. Actually, it must be a function of space-time derivatives only, being in Lorentz-invariant case a function of $\Box$. In the standard case, $\hat{T}=1$.

In these diagrams,  depicted in Fig.~1,  the double line is for the background alternating $\Psi$ and $\bar{\Psi}$ fields.
The supergraph of such structure with $2n$
legs represents itself as a ring containing $n$ links of the form depicted in Fig.~2.

\vspace*{2mm}

\begin{figure}[!h]
\begin{center}
\includegraphics[angle=0,scale=1.00]{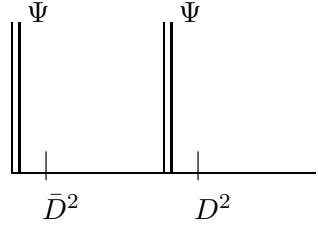}
\end{center}
\caption{The link composed by superfield propagators and background fields. Repeating of these links forms the one-loop graphs.}
\end{figure}

\vspace*{2mm}

In this case the calculations are the same as in \cite{YMEP}, so, we merely quote the result. After a subtraction of the divergence, we obtain the one-loop corrected  K\"ahlerian potential:
\bea
K^{(1)}=-\frac{1}{32\pi^2}\Psi\bar{\Psi}\ln\frac{\Psi\bar{\Psi}}{\mu^2}\, .
\eea
Adding this contribution to the classical action (\ref{s}), we get the following one-loop corrected low-energy effective action:
\bea 
\Gamma=S+\int d^8z K^{(1)}\, . \label{oneloopaction}
\eea
Note that the one-loop chiral effective potential is zero and the auxiliary fields' effective potential contributes only to higher terms in the derivative expansion, see \cite{BK0}. Therefore the explicit form of (\ref{oneloopaction})  is
\bea
\Gamma&=&
\int d^8z \Big(\phi\bar{\phi}+\Phi\bar{\Phi}-\frac{\hbar}{32\pi^2}(M+f\phi)(M+f\bar{\phi})\ln\frac{(M+f\phi)(M+f\bar{\phi})}{\mu^2}\Big)+ 
\nonumber\\&+&
\Big(\int d^6z (\frac{1}{2}(M\Phi^2+\lambda\Phi\phi^2+f\phi\Phi^2)+\frac{g}{3!}\phi^3)+h.c.\Big).
\eea
Here we kept the Planck constant $\hbar$ explicitly.
We expand this expression in series of $M$ (remind that $M$ is large)  and obtain:
\bea
\Gamma&=&
\int d^8z \Big(\phi\bar{\phi}+\Phi\bar{\Phi}-\frac{\hbar}{32\pi^2}f^2\phi\bar{\phi}(2+\ln\frac{M^2}{\mu^2})\Big)+ 
\nonumber\\&+&
\Big(\int d^6z (\frac{1}{2}(M\Phi^2+\lambda\Phi\phi^2+f\phi\Phi^2)+\frac{g}{3!}\phi^3)+h.c.\Big)+{\cal O}(M^{-1}).
\eea
Then, we eliminate $\Phi$ by its equation of motion which in the lower order in $\frac{1}{M}$ yields $\Phi=-\frac{\lambda\phi^2}{2M}+{\cal O}({M^{-2}})$, cf. \cite{BCP}. It is sufficient to conclude that in lower (zero) order, the low-energy effective action is
\bea
\Gamma&=&
\int d^8z \Big(\phi\bar{\phi}-\frac{\hbar}{32\pi^2}f^2\phi\bar{\phi}(2+\ln\frac{M^2}{\mu^2})\Big)+ 
\Big(\int d^6z \frac{g}{3!}\phi^3+h.c.\Big)+{\cal O}({M}^{-1}).
\eea
There is a  significant correction proportional to $\ln\frac{M^2}{\mu^2}$ which was a key result of \cite{BCP}.  As mentioned in \cite{BCP}, if one adds quantum fluctuations  of the light field $\phi$, 
there would  be also an additional contribution to the effective action  (the Coleman-Weinberg potential):
$\Gamma_{min}=-\frac{1}{32\pi^2}\int d^8z g^2\phi\bar{\phi}\ln\frac{g^2\phi\bar{\phi}}{\mu^2}$, so that the whole low-energy effective action would be the sum $\Gamma+\Gamma_{min}$.  If one fixes   the normalization parameter $\mu$ to be  $\mu=\alpha M$, this  will imply that the contribution of the one-loop action increases with growing of $M$, as the one-loop  contribution is of the form 
$-\frac{1}{32\pi^2}\int d^8z g^2\phi\bar{\phi}\ln\frac{g^2\phi\bar{\phi}}{M^2}$. 
We therefore conclude that neither fixing the renormalization scale $\mu$  to be of the order of $M$ nor leaving it arbitrary will avoid the appearance of large quantum corrections in the one-loop corrected effective action. 

It was demonstrated in \cite{BCP} that these  large quantum  corrections take place in a wide class of theories involving fields with very large masses. We note that suggesting that the heavy field $\Phi$ has a nontrivial background part will not essentially modify the situation since in any case the solution for $\Phi$ is proportional at least to ${M^{-1}}$, and hence all terms involving $\Phi$ in the effective action will be suppressed by $M^{-1}$, cf.  \cite{BCP}. Thus, the presence of large quantum-corrections is an universal effect in renormalizable quantum field theories with  heavy fields.
Note that the presence of the factors proportional to  $\ln\frac{M^2}{\mu^2}$  imply that this effect occurs in renormalizable theories, where the scale $\mu^2$ signifies the renormalization  scale. 

A natural question arises whether there is a way to get large quantum correction in finite theories,  where the effective action cannot depend on an arbitrary renormalization scale $\mu$. We will consider this in the subsequent section.

\section{Higher-derivative theories and Heavy States}

In this section, we discuss the decoupling effects of heavy states manifests due to  supersymmetric higher-derivative terms. We consider the following three prototypical examples. First,  we study a minimal model where derivative term is  purely of the  higher-derivative form,  and the usual two-derivative kinetic energy term is absent; this is only a ``warm-up'' toy model to  study  general impacts in the presence of higher derivatives.  Second, we study  a non-minimal model involving both the higher-derivative terms, along with  the standard two-derivative kinetic energy term.  This is considered as a subsector of an effective theory, arising from an ultraviolet complete one, such a superstring theory.  Third, we study examples of  supersymmetric higher derivative terms that break  Lorentz invariance.

\subsection{Minimal case}

We start this section with a minimal case in order  to give a simplest illustration how large quantum corrections can arise for higher-derivative theories. Of course, the model is only a toy example, as the heavy fields do not have a standard two-derivative kinetic energy term.

The corresponding theory is a higher-derivative $N=1$ supersymmetric  model with light and heavy superfields, given by
\bea
\label{s1}
S=\int d^8z (\phi\frac{\Box}{\Lambda^2}\bar{\phi}+\Phi\frac{\Box}{\Lambda^2}\bar{\Phi})+ \Big(\int d^6z (\frac{1}{2}(M\Phi^2+\lambda\Phi\phi^2+f\phi\Phi^2)+\frac{g}{3!}\phi^3)+h.c.\Big),
\eea
which is a simple generalization of (\ref{s}).
A similar theory of only one chiral superfield has been studied in \cite{ourhigh,Ant}. Here, however, we introduce a mass dimension one scale parameter   $\Lambda$ 
which enforces that  the components of the chiral superfields  have a correct dimension; effectively, $\Lambda$ plays a role of the energy scale at which the higher derivatives become important, cf.  \cite{Trodden}. In string theory compactifications this scale is naturally $\Lambda ={\cal O}(M_{Planck})$. 

Again, to simplify the study, we suggest $\phi$ to be purely external and $\Phi$ purely quantum. (In full analogy with the previous section, we can argue that if we suggest $\Phi$ to have a nontrivial background part, the situation will not be  essentially different since already the lowest approximation for $\Phi$ is proportional to ${M}^{-1}$.) The quadratic action of $\Phi$ is of the form:
\bea
\label{sq}
S_q=\int d^8z \Phi\frac{\Box}{\Lambda^2}\bar{\Phi}+ \Big(\frac{1}{2}\int d^6z (M+f\phi)\Phi^2+h.c.\Big).
\eea
This quantum action is similar to the one considered in \cite{ourhigh}. We can introduce again the superfield $\Psi=M+f\phi$. The propagator can be obtained  by a straightforward generalization of the usual Wess-Zumino case \cite{BK0}  and has the form
\bea
<\Phi(z_1)\bar{\Phi}(z_2)>=\frac{\Lambda^2}{\Box^2}\delta^8(z_1-z_2),
\eea
with again, as in the Wess-Zumino model, the chiral vertex carries the factor $-\frac{\bar{D}^2}{4}$ and the antichiral one, the factor $-\frac{D^2}{4}$.  Following  \cite{BK0}, we can calculate the superficial degree of divergence for this theory: 
\bea
\omega=2-2P-P_c-E_c,
\eea
where $P$ is a number of all propagators in the theory, $P_c$ is a number of chiral propagators, i.e., $<\Phi\Phi>$ and $<\bar{\Phi}\bar{\Phi}>$ propagators only, $E_c$ is a number of external chiral lines.  (It is easy to show that  the nonminimal theory we are considering possesses  the same superficial degree of divergence, since the Wess-Zumino kinetic term yields only subleading contributions to propagators of $\Phi$, $\bar{\Phi}$.)

It is clear that for any $E_c\geq 1$ and $P\geq 1$, one has $\omega<0$, so, the theory is ultraviolet finite. At the same time, in the usual Wess-Zumino model one has $\omega=2-P_c-E_c$, so, if the Feynman supergraph involves no chiral propagators but only $<\Phi\bar{\Phi}>$ propagators, it can yield divergent corrections to the kinetic term, while in our case there is no divergent corrections at all. We emphasize that within this study we treat $\Lambda$ as a finite parameter of the (effective) field theory, thus there is no need to  introduce counterterms as  $\Lambda\to\infty$.

The low-energy effective action presented by a sum over the supergraphs depicted in Fig.~1 is
\bea
\label{g1}
\Gamma_1=\int d^8z\sum\limits_{n=1}^{\infty}\frac{1}{2n}\left[\frac{\Psi\bar{\Psi}\Lambda^4}{\Box^4}\frac{\bar{D}^2D^2}{16}\right]^n
\delta^8(z-z')|_{z=z'}.
\eea
This sum is analogous to that one used in the usual Wess-Zumino and SYM cases \cite{YMEP}, with the only difference {\bf being} the fact that we have $\Box^2$ in the denominator instead of the usual $\Box$. To obtain the low-energy (K\"{a}hlerian) contribution to the effective action, we disregard all terms where derivatives act on  background fields.
In order to sum contributions,  we use the property of the projection operator $\left(\frac{\bar{D}^2D^2}{16\Box}\right)^n=\frac{\bar{D}^2D^2}{16\Box}$. Subsequently,  we employ  ``the shrinking of the loop to a point'' via the well-known identity $\frac{\bar{D}^2D^2}{16}\delta^8(z-z')|_{z=z'}=1$, and finally,  employ the sum $\sum\limits_{n=1}^{\infty} a^n=-\ln(1-a)$. 
We arrive at the following  result for the K\"{a}hlerian effective potential, analogous to  that in \cite{ourhigh}:
\bea
\label{g2}
K^{(1)}=\frac{1}{2}\int\frac{d^4k}{(2\pi)^4}\frac{1}{k^2}\ln\Big[1+\frac{|\Psi|^2\Lambda^4}{k^6}\Big],
\eea
which, by removing the field independent part, takes the form
\bea
K^{(1)}=\frac{1}{2}\int\frac{d^4k}{(2\pi)^4}\frac{1}{k^2}\ln\Big[\frac{k^6}{\Lambda^4}+|\Psi|^2\Big],
\eea
which yields
\bea
\label{res}
K^{(1)}=\frac{c_0}{32\pi^2}(\Psi\bar{\Psi}\Lambda^4)^{1/3},
\eea
where $c_0$ is a finite dimensionless constant whose value can be found in \cite{ourhigh}.
We can again expand the low-energy effective action in series of  $M$ (recall that $\Psi=M+f\phi$). As a result, we obtain
\bea
\label{res1}
K^{(1)}=\frac{c_0}{288\pi^2}\frac{\Lambda^{4/3}}{M^{4/3}}f^2 \phi\bar{\phi}+{\cal O}(\frac{\Lambda^{4/3}}{M^{7/3}}).
\eea
It is clear that after the rescaling $\phi(1+\frac{c_0}{288\pi^2}\frac{\Lambda^{4/3}}{M^{4/3}}f^2)^{1/2}\to\phi$, together with the rescalings of constant parameters, the one-loop corrected effective action $\Gamma=S+\int d^8 z K^{(1)}$ reproduces the classical action of the theory involving light superfields only. Thus, the decoupling theorem is formally satisfied. However, the key point in interpreting this result  relies on the magnitude of the scale $\Lambda$. We see that if $\Lambda\ll M$, the perturbative contribution to the effective action is strongly suppressed, as $\left(\frac{\Lambda}{M}\right)^{-4/3}$. However, for  $\Lambda\simeq M$,  when there is only one energy scale in the theory, the quantum correction becomes compatible with the tree-level effective action.  The case  $\Lambda>M$  has a natural occurrence  in string theory, where
$\Lambda={\cal O}(M_{Planck})$ and $M<M_{Planck}$, e.g., in perturbative heterotic string theory 
$M\simeq M_{String}=10^{-2}M_{Planck}$). In this case  quantum contributions begin to dominate. 
 
 We also note that the mechanism of  large quantum corrections described in section 2 cannot be applied to the case with finite quantum contributions, discussed in this section, since now  quantum corrections  do not depend on the arbitrary normalization parameter $\mu$ which can arise only as a consequence of subtractions of divergences. 
In the higher-derivative field theory models, instead of $\mu$,  there is another natural energy scale $\Lambda$, which describes a characteristic energy scale at which the higher derivatives become important.  However, despite of essentially different mechanisms for  the quantum corrections in divergent and higher-derivative finite theories, the general structure of quantum corrections turns out to be rather similar in both cases;   the logarithmic quantum corrections,  proportional to  $\mu$, in  renormalizable theories, and   to $\Lambda$, in finite higher derivative  theories,  is very analogous. 

We can generalize these studies to a generic theory whose action of quantum fields is given by (\ref{gentheory}), instead of (\ref{sq}).  In this case, $\hat{T}$ is a generic scalar operator commuting with supercovariant derivatives. In particular, we can have the case $\hat{T}=1$ which corresponds to the usual Wess-Zumino model. Indeed, we would have the same background fields $\Psi$ and $\bar{\Psi}$, but our propagator $<\Phi\bar{\Phi}>$ will be of the form:
\bea
<\Phi(z_1)\bar{\Phi}(z_2)>=\frac{1}{\hat{T}\Box}\delta^8(z_1-z_2),
\eea
so, we can adapt the results given by (\ref{g1},\ref{g2}) and find
\bea
K^{(1)}=\frac{1}{2}\int\frac{d^4k}{(2\pi)^4}\frac{1}{k^2}\ln\Big[T^2(k)k^2+|\Psi|^2\Big],
\eea
where $T(k)$ is a Fourier transform  of the operator $\hat{T}$. In the simplest case $T(k)=\frac{k^{2n}}{\Lambda^{2n}}$ ($n\geq 1$), where $\Lambda$ again plays a role of an energy scale at which the higher derivatives dominate, it is easy to find this integral. The result is
\bea
K^{(1)}=\frac{1}{32\pi^2}\Gamma(\frac{1}{2n+1})\Gamma(\frac{2n}{2n+1})\left(\frac{|\Psi|^2}{\Lambda^2}\right)^{\frac{1}{2n+1}}.
\eea
Again, defining  $\Psi=M+\lambda\phi$ and expanding  this expression as a power series in $\lambda$, we obtain
\bea
K^{(1)}=k_0(\Lambda^{4n}M^2)^{\frac{1}{2n+1}}\Big[\frac{\lambda^2\phi\bar{\phi}}{M^2}+{\cal O}({M^{-3}})
\Big],
\eea
where $k_0$ is a purely numerical constant which does not depend on any physical scale.
Therefore the scale of this expression is completely characterized by  $\left(\frac{\Lambda}{M}\right)^{4n/(2n+1)}$. 

If we take  $\Lambda\simeq M$, namely if the  theory involves only one characteristic energy scale, we have
\bea
K^{(1)}=\frac{k_0}{(2n+1)^2}\lambda^2\phi\bar{\phi}+{\cal O}({M^{-1}}).
\eea
Then, the quantum correction caused by coupling of a light superfield $\phi$ with heavy ones is not suppressed, as in the previous example, although we again can argue that this result is formally consistent with the decoupling theorem.
And if we suggest that $\Lambda\gg M$, as e.g. in the above-mentioned case, when $M\simeq M_{String}\simeq 10^{-2}\Lambda$, this correction begin to dominate. Namely,  for  $\Lambda=\gamma M$, with $\gamma \gg 1$, we have
\bea
K^{(1)}=\gamma^{\frac{4n}{2n+1}}\frac{k_0}{(2n+1)^2}\lambda^2\phi\bar{\phi}+{\cal O}({M^{-1}}).
\eea
Thus,  the quantum correction is large since $\gamma^{\frac{4n}{2n+1}}\simeq 10^4$.

\subsection{Nonmimimal case}

Another example of the higher-derivative superfield theory, that includes the standard kinetic energy terms, {\bf has been} discussed in \cite{Ant,ourhigh}. The action is of the form:
\bea
\label{s12}
S&=&\int d^8z \Big(\phi\bar{\phi}+\Phi(1+\frac{\Box}{\Lambda^2})\bar{\Phi}\Big)+ \nonumber\\ &+&
\Big(\int d^6z (\frac{1}{2}(M\Phi^2+\lambda\Phi\phi^2+f\phi\Phi^2)+\frac{g}{3!}\phi^3)+h.c.\Big)\, .
\eea
For simplicity we take $\phi$ chiral superfield to be massless.
Again,  we choose  $\phi$ to be a purely external one, and $\Phi$ to be a purely internal one. (For the sake of simplicity we do not introduce  higher derivatives for the light superfields. In the effective theory of light fields, only, these terms are suppressed by  ${\cal O}(\Lambda^{-1})$   and decouple.)
We carry out a summation over the supergraphs depicted in Fig.~1. The  result, after the Wick rotation, is {\bf given by}
\bea
\label{intres}
K^{(1)}=\frac{1}{2}\int\frac{d^4k}{(2\pi)^4}\frac{1}{k^2}\ln\Big[k^2(1+\frac{k^2}{\Lambda^2})^2+|\Psi|^2
\Big].
\eea
The explicit result for this expression is  cumbersome. Nevertheless, we can proceed with this integral in some characteristic cases.

We proceed  by considering the object $J=\frac{dK^{(1)}}{d(\Psi\bar{\Psi})}$. A replacement $d^4k\to \pi^2 t dt$, with $t=k^2$, yields
\bea
\label{fract}
J=\frac{1}{32\pi^2}\int_0^\infty dt
\frac{\Lambda^4}{t(t+\Lambda^2)^2+\Lambda^4|\Psi|^2}=\frac{1}{32\pi^2}\int_0^\infty dt
\frac{\Lambda^4}{(t+A)(t+B)(t+C)},\label{J}
\eea
where $A,B,C$ are three roots of the denominator, taken with opposite signs, i.e., $t(t+\Lambda^2)^2+\Lambda^4|\Psi|^2=(t+A)(t+B)(t+C)$. 
Then, we can write
\bea
\frac{1}{t(t+\Lambda^2)^2+\Lambda^4|\Psi|^2}=\frac{1}{Q}\left[\frac{B-C}{t+A}+\frac{C-A}{t+B}+\frac{A-B}{t+C}\right],\label{propagator}
\eea
where $Q=AB(A-B)+BC(B-C)+CA(C-A)$. It is clear therefore that at least one of the numbers $A-B$, $B-C$, $C-A$ will be negative, and hence at least one of the  residua of the propagator  (\ref{propagator}) will be negative.  Thus the  action  (\ref{s12}) unavoidably involves ghosts.

We should however note  that a fundamental, ultraviolet complete theory, such as string theory, should be ghost-free. Therefore, we can treat our result in the following manners. First, one can suppose that the specific higher-derivative terms also represent themselves as contributions in the effective theory where the higher derivatives arise as a consequence of an integration over additional matter fields as it occurs, e.g., in gravity theories coupled to additional matter fields   \cite{AM}.
Second, we can treat higher-derivative terms as a next approximation in a derivative expansion of a fundamental non-local theory where the ghosts are avoided \cite{Modesto}. We note that if we abandon the Lorentz invariance, the higher derivatives can be introduced in a unitary manner through appropriate contractions with  Lorentz-breaking vectors or tensors \cite{CMR}. We consider this situation later.
 
The straightforward integration of (\ref{J})  allows one to  obtain the explicit result in terms   of  roots $A,B,C$:
\bea
J=\frac{\Lambda^4}{16\pi^2Q}\Big[A\ln\frac{B}{C}+\ln\frac{C}{A}+C\ln\frac{A}{B}\Big].
\eea
In principle, one can use here the explicit expressions for $A,B,C$.  They are given in the Appendix, along with their 
explicit asymptotic behaviour in  $\frac{|\Psi|^2}{\Lambda^2}\ll 1$ and $\frac{|\Psi|^2}{\Lambda^2}\gg 1$ regimes. 
The explicit form of  (\ref{J}) in these regimes can be also analysed directly, by introducing  the dimensionless quantity $R^2=\frac{|\Psi|^2}{\Lambda^2}$, i.e., by writing (\ref{J}) as
\bea
\label{jred}
J=\frac{1}{32\pi^2}\int_0^\infty du\frac{1}{u(u+1)^2+R^2}\, , 
\eea
and find the asymptotic behaviour of this integral in  $R\ll 1$ and $R\gg 1$ regimes.
 The case $R\ll 1$  effectively 
corresponds to 
$M\ll \Lambda$ and results in the expansion:
\bea
J|_{R\to 0}=-\frac{1}{32\pi^2}[\ln R^2+1+R^2(4\ln R^2+\frac{19}{3})]+{\cal O}(R^3).
\eea 
The  case $R\gg 1$  corresponds to $M\gg\Lambda$ and results in
\bea
J|_{R\to\infty}=-\frac{1}{96\pi^2}\frac{\ln R^2}{R^{4/3}}+{\cal O}(R^{-2}).
\eea
Thus, by integrating these expressions, respectively, we obtain
\bea
K^{(1)}_{R\to 0}&=&-\frac{1}{32\pi^2}(|\Psi|^2\ln\frac{|\Psi|^2}{\Lambda^2}+
2\frac{|\Psi|^4}{\Lambda^2}\ln\frac{|\Psi|^2}{\Lambda^2}+\frac{13}{6}\frac{|\Psi|^4}{\Lambda^2})+\cdots;\nonumber\\
K^{(1)}_{R\to\infty}&=&-\frac{1}{32\pi^2}|\Psi|^{2/3}\Lambda^{4/3}(\ln\frac{|\Psi|^2}{\Lambda^2}+\frac{3}{2})+\cdots,
\eea
where again $\Psi=M+f\phi$.  Further expansion of  the above expressions in terms of  the small field $\phi\ll M$, and disregarding  field independent terms 
(whose contribution to the effective action vanishes because of properties of the integral over Grassmannian variables), 
yields the following  expressions:
\bea
K^{(1)}_{M\ll\Lambda}&=&-\frac{1}{32\pi^2}f^2\phi\bar{\phi}\ln\frac{M^2}{\Lambda^2}+\cdots;\nonumber\\
K^{(1)}_{M\gg \Lambda}&=&-\frac{1}{288\pi^2}(\frac{\Lambda}{M})^{4/3}f^2\phi\bar{\phi}\ln\frac{M^2}{\Lambda^2}+\cdots.
\eea

Note,  that in the second case $M\gg \Lambda$ the correction is suppressed.  It is however the first case  
$M\ll \Lambda$ that one encounters  in typical string theory compactifications. In particular, in perturbative heterotic string compactification $\Lambda = {\cal O}(M_{Planck})$ and  $M={\cal O}(M_{String})$, where 
 $M_{String}=10^{-2}M_{Planck}$.  Note that  in this case there are  large logarithmic corrections.

\subsection{Lorentz-breaking case}

In this subsection we  address the study of Lorentz violating terms due to heavy superfields.  As a first step we introduce  supersymmetric higher derivative terms that break Lorentz invariance.  For specific choices of the Lorentz breaking vector in  the higher-derivative terms, as proposed in \cite{MP},  the theory can still maintain unitarity.  As a prototype, we consider the following extension of the Wess-Zumino model:
\bea
S&=&\int d^8z \Big[\bar{\Phi}(1-\frac{1}{\Lambda^2} (n\cdot\pa)^2)\Phi+\bar{\phi}\phi\Big]+\nonumber\\&+& \Big(\int d^6z (\frac{1}{2}(M\Phi^2+\lambda\Phi\phi^2+f\phi\Phi^2)+\frac{g}{3!}\phi^3)+h.c.\Big).
\eea
Here, $n_a$ is a dimensionless Lorentz-breaking vector which in principle can be chosen to be space-like ($n_an^a=1$), to avoid higher time derivatives which spoil the unitarity.  (In reality, we do all calculations in the Euclidean space, and hence this problem is avoided). This is a Myers-Pospelov-like extension  to the  case of the Wess-Zumino model.  
Here we introduced $(n\cdot \pa)^2$ to simplify the integration. We note that the  introduction of the higher-derivative Lorentz-breaking  terms in this manner, i.e. through terms $(n\cdot \pa)^N$ with different values of $N$, was discussed in \cite{MP}.

Again, we consider the same type of graphs as in Fig.~1. We repeat the calculations, given by (\ref{g1}--\ref{g2}), along the lines discussed earlier.  Actually, the only difference is in replacement of $\frac{\Lambda^2}{\Box^2}$ by $\left[\Box(1-\frac{1}{\Lambda^2} (n\cdot\pa)^2)\right]^{-1}$. Repeating the summation, we arrive at
\bea
\label{intres1}
K^{(1)}=\frac{1}{2}\int\frac{d^4k}{(2\pi)^4}\frac{1}{k^2}\ln\Big[k^2\left(1+\frac{(n\cdot k)^2}{\Lambda^2}\right)^2+|\Psi|^2
\Big].
\eea
We make  a replacement $k_ak_b=\frac{1}{4}\eta_{ab}k^2$, which is valid within any integral over momenta of the form $\int\frac{d^dk}{(2\pi)^d}f(k)k_{a_1}k_{a_2}\ldots k_{2m}$. For example, in the simplest case $m=1$, with arbitrary values of $d$ and $N$, one has $$\int\frac{d^dk}{(2\pi)^d}\frac{k_ak_b}{(k^2+M^2)^N}=\frac{1}{d}\eta_{ab}\int\frac{d^dk}{(2\pi)^d}\frac{k^2}{(k^2+M^2)^N}=
\frac{1}{2(4\pi)^{d/2}}(M^2)^{d/2+1-N}\frac{\Gamma(N-1-d/2)}{\Gamma(N)};$$
and for larger even $m$ this formula can be naturally generalized. (We note that since the effective potential is a scalar, the Lorentz-breaking vector $n_a$ can enter only through a contraction $n^an_a$ which is equal to 1; moreover, since our effective potential is finite, it has no singularities,  and therefore its behaviour will not  be modified qualitatively by  changing the vector $n_a$.)
 
We proceed,  as in the previous section, by calculating the asymptotic  form of the exact integrals in   $|\Psi|^2\ll \Lambda^2$ (effectively, $M\ll \Lambda$)  and in  $|\Psi|^2\gg\Lambda^2$ (effectively, $M\gg\Lambda$) regime. The result is:
for $|\Psi|^2\ll \Lambda^2$,
\bea
K^{(1)} = -\frac{13 \Psi ^4}{768 \pi ^2 \Lambda ^2}-\frac{\Psi ^2 \ln \left(\frac{\Psi ^2}{4 \Lambda ^2}\right)}{32 \pi ^2}-\frac{\Psi ^4 \ln \left(\frac{\Psi ^2}{4 \Lambda ^2}\right)}{64 \pi ^2 \Lambda ^2} +\cdots,
\eea
and for $|\Psi|^2\gg \Lambda^2$,
\bea
K^{(1)} = \frac{\Psi^2 \left(\frac{\Lambda ^2}{\Psi^2}\right)^{2/3}}{4\ 2^{2/3} \sqrt{3} \pi }-\frac{\Lambda^2 \ln \left(\frac{\Psi^2}{4 \Lambda ^2}\right)}{12 \pi^2}+\cdots.
\eea
Expanding the above expressions in power series for the small $\phi$, we have
\bea
K^{(1)}_{M\ll \Lambda}=-\frac{\lambda^2\phi\bar{\phi}}{32\pi^2}[3+\ln\frac{M^2}{4\Lambda^2}]+{\cal O}(\frac{M}{\Lambda},\frac{1}{M})+\cdots,
\eea
and 
\bea
K^{(1)}_{M\gg \Lambda}=(\frac{\Lambda^2}{M^2})^{2/3}\frac{1}{2^{2/3}3^{5/2}\pi}\lambda^2\phi\bar{\phi}+{\cal O}(\frac{\Lambda}{M},\frac{1}{M})+\cdots,
\eea
{\bf respectively.} We see again that in the second case  corrections are suppressed. It is however the first case with $\Lambda\gg M$ that one encounters in the string theory compactification. In this case the corrections have logarithmic enhancement. 

It is instructive to compare our results with another type  of Lorentz-violating supersymmetric theories, discussed in \cite{LVWZ}, where the classical action of the form
\bea
\label{sfigeral}
S=\int d^8z\Phi(1+\rho\Delta^{z-1})\bar{\Phi}+(\int d^6z W(\Phi) +h.c.)
\eea
was considered, with $W(\Phi)=\frac{m}{2}\Phi^2+\frac{\lambda}{3!}\Phi^3$. 
There, for the same $\Psi=m+\lambda\phi$ (where, however, the mass $m$ was not enforced to be large as in our case), the result was
\bea
\label{reshor}
K^{(1)}=c\rho^{-1/z}(\Psi\bar{\Psi})^{1/z},
\eea
where $z$ is a critical exponent, $\rho$ is a small constant whose mass dimension is $-2(z-1)$, and $c$ is a some number of the order of 1. We can modify this theory within the framework of our approach, so that the action would be
\bea
\label{s12a}
S&=&\int d^8z \Big(\phi\bar{\phi}+\Phi(1+\rho\Delta^{z-1})\bar{\Phi}\Big)+ \nonumber\\ &+&
\Big(\int d^6z (\frac{1}{2}(M\Phi^2+\lambda\Phi\phi^2+f\phi\Phi^2)+\frac{g}{3!}\phi^3)+h.c.\Big).
\eea
In this case, one has $\Psi=M+f\phi$. Repeating the calculations, 
and introducing a  cut-off scale $\Lambda$ so that we can estimate $\rho=\frac{\alpha}{\Lambda^{2z-2}}$ by  dimensional analysis, with $\alpha$ being a number of the order of 1, we arrive at the result (\ref{reshor}), which, for our choice of $\Psi$, can be expanded as
\bea
K^{(1)}=c\frac{\alpha^{-1/z}}{z^2}(\frac{M}{\Lambda})^{\frac{2}{z}-2}f^2\phi\bar{\phi}+\cdots,
\eea
where dots are for higher orders in expansion in ${M}^{-1}$.
(All the dependence on $\Lambda$ is concentrated in the factor $(\frac{M}{\Lambda})^{\frac{2}{z}-2}$, so, dots are for essentially smaller terms.) 
We see that for  $\Lambda$ and $M$  of the same order, this correction is not suppressed. Moreover, since typically $z>1$, and $\frac{2}{z}-2<0$, we see that for  $M<\Lambda$ (which is a natural  regime as  already argued)
the low-energy effective action increases as  $M$ grows. Moreover, in this case the factor $(\frac{M}{\Lambda})^{\frac{2}{z}-2}$ is large, tending to $10^4$ for a large $z$.

\section{Summary}

In this paper we discussed the problem of decoupling of heavy states in four-dimensional $N=1$ supersymmetric field theory  sector with light and heavy chiral superfields.
We encountered an essential difference between the  renormalizable theories, exhibiting divergent quantum contributions, and effective theories with higher derivative terms which have finite quantum corrections. In renormalizable quantum field theories,  after renormalization,  quantum corrections at the $L$-loop level one can have contributions proportional to $|\phi|^2\left(\ln\frac{|\phi|^2}{\mu^2}\right)^L$, where,  $\mu$ is the renormalization scale. Since the only dimensional scale in the theory is a large mass $M$, it is natural to fix $\mu=M$, so, one will have significant contributions which, moreover, increase as $M$ grows.  Of course, by changing the $\mu$ to be a low energy scale, this would then result in renormalized couplings,  in front of such terms, which are defined at a low energy scale $\mu$.

In effective theories with higher derivative terms the quantum corrections are finite 
and  thus do not involve the renormalization scale.
However, if one considers higher-derivative extensions, one introduces the scale $\Lambda$ characterizing the energy at which the higher-derivative terms become important \cite{Trodden}. As we have shown, if this scale is large enough, i.e. of the order of the mass of the heavy superfield, the quantum corrections  due to these terms  also become significant.
Therefore, we found that instead of the mechanism based on the renormalization  the effects of higher derivative terms 
result in large finite quantum corrections. 
This effect is realized for both Lorentz-invariant and  Lorentz-breaking  higher derivative examples. 
In principle, this result can be formally understood if we treat  higher-derivative terms as a type of a higher-derivative ``regularization''  that ensures  finiteness of  quantum corrections. It is  therefore clear that in the absence of the regularization the one-loop effective action will diverge. 
However, within our approach the higher-derivative terms are treated not as a regularization but are  a fundamental ingredient of the effective theory. Therefore, we have large quantum corrections which, following \cite{CMR}, can be interpreted as a  sign of a fine-tuning in a corresponding effective theory of light fields only. Furthermore, for an example of an effective theory of the perturbative heterotic string compactification, $M\simeq 10^{-2}\Lambda$,  and thus large quantum corrections, proportional to $\ln(\frac{M^2}{\Lambda^2})$ are unavoidable. These results also imply,  that  theories with  heavy superfields, both those with higher derivatives and those without them,  possess
quantum corrections  that can significantly modify the low-energy  effective theory of light fields. 
We also expect that the presence of the fields with a large mass could have  cosmological impact
since  the presence of large quantum corrections would strongly modify the observable values of physical variables. Also, since the galileon models naturally involve higher derivatives \cite{Rattazzi}, it is natural to expect that studies of the higher-derivative field theory models, in particular, the those with finite quantum corrections, can be relevant within the context of galileon cosmology.

A generalization of this study to  would involve higher derivative couplings of chiral superfields  to gauge superfields. However, up to now, the only models studied in this manner involve higher derivatives in the  gauge sector \cite{higherQED}. 
Another direction for future studies should involve a more general and systematic study of  Lorentz-breaking supersymmetric theories, in particular more generic models for chiral matter, e.g., Lorentz-breaking extensions of models discussed in \cite{BCP}, as well as superfield analogues of the models discussed in \cite{MP}.

\vspace*{1cm}
{\noindent {\bf \large Acknowledgments}}

\vspace*{2mm}
This work was partially supported by Conselho Nacional de
Desenvolvimento Cient\'\i fico e Tecnol\'ogico (CNPq). A.Yu.P. has
been supported by the CNPq project No. 303438-2012/6.  M.C. 
is supported in part by the  DOE (HEP) Award DE-SC0013528, the 
Fay R. and Eugene L. Langberg Endowed Chair (M.C.) and the Slovenian 
Research Agency (ARRS). M.C. and  A.Yu.P. would like to thank the organizers of the 
workshop: `` Theoretical Frontiers in Black Holes  and Cosmology,''  in Natal, Brazil, where the work was initiated, for hospitality.

\vspace*{5mm}

{\centerline{\bf APPENDIX}}

\vspace*{0.5cm }

Here we give explicit expressions for the roots  $A,B,C$ of the denominator of (\ref{fract}). Using the Cardano formula, we can write

\bea
A&=&\frac{1}{3}(2\Lambda^2+Q+\bar{Q}),\nonumber\\
B&=&\frac{1}{3}\left(2\Lambda^2+\left(\frac{-1+i\sqrt{3}}{2}\right)Q+\left(\frac{-1-i\sqrt{3}}{2}\right)\bar{Q}\right),\nonumber\\
C&=&\frac{1}{3}\left(2\Lambda^2+\left(\frac{-1-i\sqrt{3}}{2}\right)Q+\left(\frac{-1+i\sqrt{3}}{2}\right)\bar{Q}\right),
\eea
where
\bea
Q&=&\sqrt[3]{\frac{27\Lambda^4|\Psi|^2-2\Lambda^6+\sqrt{729\Lambda^8|\Psi|^4-108\Lambda^{10}|\Psi|^2}}{2}},\nonumber\\
\bar{Q}&=&\sqrt[3]{\frac{27\Lambda^4|\Psi|^2-2\Lambda^6-\sqrt{729\Lambda^8|\Psi|^4-108\Lambda^{10}|\Psi|^2}}{2}}.
\eea

It remains only to find these roots  in the approximation:   $\Lambda^2\ll|\Psi|^2$ (i.e. $R\equiv \frac{|\Psi|}{\Lambda}\gg 1$) and for $\Lambda^2\gg|\Psi|^2$ (i.e. $R\ll 1$).
It is more convenient to consider the dimensionless objects $\tilde{A}\equiv\frac{A}{\Lambda^2}$, $\tilde{B}\equiv\frac{B}{\Lambda^2}$, $\tilde{C}\equiv\frac{C}{\Lambda^2}$. Actually, $\tilde{A}$, $\tilde{B}$, $\tilde{C}$ are the roots of the denominator of the Eq. (\ref{jred}) multiplied by $-1$. (This factor is needed to return to the form of the roots corresponding to the Minkowski space.) 

For  $R\ll 1$, we find
\bea
\tilde{A}&=&{\cal O}(R^2); \nonumber\\
\tilde{B}&=&1+{\cal O}(R);\nonumber\\
\tilde{C}&=&1+{\cal O}(R).
\eea
For  $R\gg 1$, we have
\bea
\tilde{A}&=&R^{2/3}+\frac{2}{3}+{\cal O}(R^{-1/3}); \nonumber\\
\tilde{B}&=&R^{2/3}+\frac{2}{3}+{\cal O}(R^{-1/3});\nonumber\\
\tilde{C}&=&R^{2/3}+\frac{2}{3}+{\cal O}(R^{-1/3}).
\eea
We see that the roots are positive, so, there are no tachyons in the theory, however, as we already mentioned, the ghosts are unavoidable.

\end{document}